\documentclass{article}
\usepackage{spconf,amsmath,graphicx}
\usepackage{amssymb}
\usepackage{subfig}
\usepackage{float}
\usepackage{cite}
\usepackage{diagbox}
\usepackage{tabularx,booktabs}
\usepackage{cellspace}
\usepackage{color}
\usepackage{caption}
\usepackage{varwidth}
\usepackage{afterpage}

\title{PRIVACY-AWARE COMMUNICATION OVER A WIRETAP CHANNEL \\ WITH GENERATIVE NETWORKS}
\name{Ecenaz Erdemir, Pier Luigi Dragotti and Deniz G\"{u}nd\"{u}z 
}
\address{Department of Electrical and Electronic Engineering, Imperial College London, UK}
\begin{document}
%
\maketitle
\begin{abstract}
\vspace{-0.15cm}
We study privacy-aware communication over a wiretap channel using end-to-end learning. Alice wants to transmit a source signal to Bob over a binary symmetric channel, while passive eavesdropper Eve tries to infer some sensitive attribute of Alice's source based on its overheard signal. Since we usually do not have access to true distributions, we propose a data-driven approach using variational autoencoder (VAE)-based joint source channel coding (JSCC). We show through simulations with the colored MNIST dataset that our approach provides high reconstruction quality at the receiver while confusing the eavesdropper about the latent sensitive attribute, which consists of the color and thickness of the digits. Finally, we consider a parallel-channel scenario, and show that our approach arranges the information transmission such that the channels with higher noise levels at the eavesdropper carry the sensitive information, while the non-sensitive information is transmitted over more vulnerable channels.

\end{abstract}

\vspace{-0.7cm}
\section{INTRODUCTION}
\vspace{-0.3cm}
Secrecy and privacy in data communication and data sharing systems have been extensively studied in the literature \cite{shannon1949communication, wyner1975wire,PhysLayer,statInf, BizBook, SkoglundEpsPriv, Funnel,TIFS, FunnelLimits, Rassouli:TIFS:20, APUT_HTAdv, Hameed:TIFS:21}. Although deep learning applications of data transmission have also been well investigated, deep learning in wireless communications and physical layer security has only recently become popular \cite{gunduz2019mlintheair, o2017dlforphylayer}.
The similarity between the communication systems and end-to-end learning motivates the use of autoencoder based neural network architectures, which simultaneously learn encoding and decoding \cite{o2017dlforphylayer, bourtsoulatze2019djscc}.
Recently, it has been shown that end-to-end approaches can also be utilized for physical layer secrecy \cite{WiretapCodeAE, Marchioro2020Adversarial, fritschek2019gaussian,Fritschek2020mine}. In a wiretap channel setting, these techniques exploit the physical characteristics of the legitimate receiver's channel over the eavesdropper's, and allow communication with secrecy guarantees.

In this work, we consider a wiretap channel scenario in which Alice wants to deliver its source, $S^m$, to Bob over a noisy communication channel, while a passive eavesdropper Eve tries to infer a latent sensitive information $T$ about $S^m$. For example, $S^m$ may be an image or a video captured by Alice, while $T$ may be the presence of a particular object or an activity within the scene. We assume binary symmetric channels (BSCs) from Alice to both Bob 
and Eve. Our aim is to optimize the trade-off between the reconstruction distortion of source $S^m$ at Bob and the privacy leakage of $T$ to Eve, which is measured by the mutual information (MI) between the sensitive information and the noisy codewords observed by Eve. Note that, the wiretap channel model considered here is normally studied in the context of secure communications. Indeed, when $T=S^m$, our problem becomes a special case of the one studied in \cite{Merhav2008}. We, instead, call this ``privacy-aware communications'' since secrecy typically focuses on making the information leakage negligible, while privacy tolerates some leakage in return of utility \cite{BlochSurvey}. Hence, we propose a privacy-utility trade-off (PUT) for communication over the wiretap channel by balancing the information leakage to the eavesdropper and the distortion at the legitimate receiver. We highlight that in the special case of identical channels to Bob and Eve, our problem also reduces to the well-known privacy funnel \cite{Funnel} with a noisy communication channel. In that scenario, Bob and Eve merge into a single receiver, to which we want to send $S^m$ with the highest fidelity while hiding $T$. Therefore, our problem generalizes both the wiretap channel and the privacy funnel problems. Additionally, unlike in \cite{Merhav2008} and \cite{Funnel}, we follow a data-driven approach by using an encoder-decoder pair, represented by a VAE network and a classifier which represents the eavesdropper. 


Similar data-driven wiretap channel approaches have recently been proposed for Gaussian channels in \cite{WiretapCodeAE, fritschek2019gaussian, Marchioro2020Adversarial, Fritschek2020mine}. However, \cite{fritschek2019gaussian, Fritschek2020mine, WiretapCodeAE} focus on channel coding, and \cite{fritschek2019gaussian, Fritschek2020mine} enforce coding structure to the encoder, while we carry out end-to-end joint learning corresponding to a JSCC approach. In addition, unlike these works, we are interested in hiding an underlying sensitive information that is correlated with, but different from the original signal. The same problem is considered in \cite{Marchioro2020Adversarial} for an additive white Gaussian channel using a generative adversarial network (GAN), which minimizes the distortion of the reconstructed signal at the legitimate receiver while characterizing the privacy with a constraint on the likelihood of the sensitive information.
On the other hand, we propose a PUT for a BSC wiretap channel using a VAE-based neural network architecture. 

VAEs provide several advantages in this framework compared to standard autoencoders (AEs) \cite{bourtsoulatze2019djscc}. They embed the input to a distribution rather than a point, and a random channel input is sampled from the latent distribution rather than being generated by the encoder directly. Hence, VAEs are more aligned with the stochastic encoding approach employed in information theoretic derivation of the wiretap channel capacity \cite{wyner1975wire, Merhav2008}. Additionally, VAEs provide significant control over how to model the latent distribution, since the encoder is designed as a generative network. This is difficult to achieve within the AE framework, and also allows a tractable calculation of the variational approximations of our cost function based on MI. Last but not least, it is challenging to optimize AEs for communication over discrete channels due to their non-differentiability, whereas sampling discrete codewords from a latent distribution is possible for VAEs. 


We apply our approach to privacy-aware image transmission and show that while the receiver can reconstruct high quality images, the eavesdropper is confused about the sensitive information. We also consider a parallel-channel case in which Bob and Eve might experience different noise levels over each channel. We show that our end-to-end approach judiciously adjusts its transmission to exploit the more secure channels to transmit the sensitive information.

\vspace{-0.45cm}
\section{PROBLEM STATEMENT}
\begin{figure}[pt]
\centering
\includegraphics[width=8.6cm]{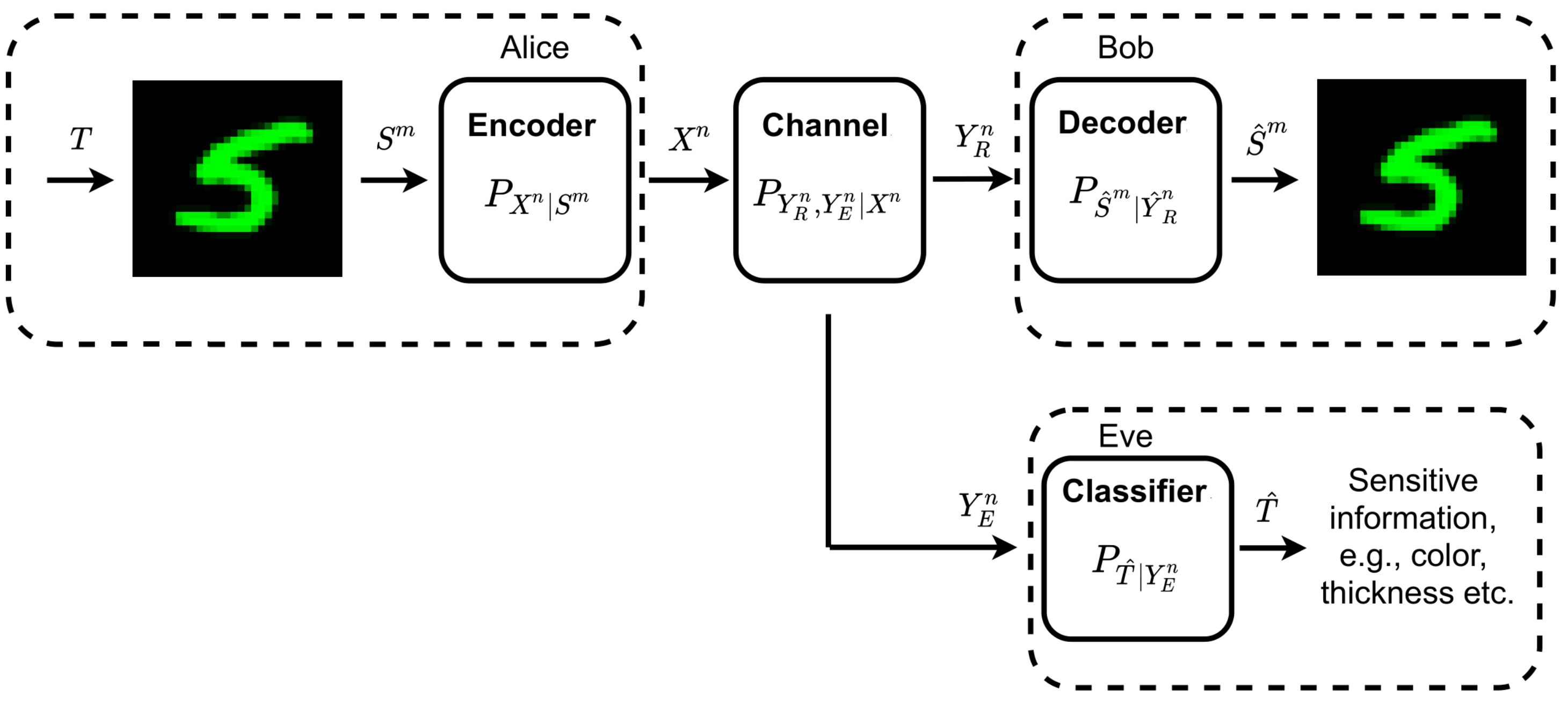}
\caption{Communication system with wiretap channel.} 
\label{fig:wiretapscenario}
\end{figure}

\vspace{-0.35cm}
We consider a communication scenario in which a user wants to reliably transmit data from one point to another over a noisy communication channel, while a passive eavesdropper tries to infer a latent sensitive information through its noisy observation of the transmitted signal. Fig. \ref{fig:wiretapscenario} illustrates the communication problem via a simple example. Alice wants to reveal her data $S^m \in \mathcal{S}$, e.g., images of the applicants for a certain job position, to Bob over a noisy channel. Eve eavesdrops through her own channel and receives a noisy version of the transmitted signal by Alice. Eve's goal is to extract Alice's sensitive information $T \in \mathcal{T}$, e.g., ethic or socioeconomic background of the applicants, which is correlated with $S^m$ but not explicitly observed by any of the involved parties. Alice's goal, on the other hand, is to encode the source such that it can be reconstructed by Bob with high fidelity, while the sensitive information $T$ cannot be accurately detected by Eve.
The source is encoded into codewords $X^n \in \{x_1, \dots, x_n\}$, where $X_i \in \mathcal{X}=\{0,1\}$, by a stochastic encoding function $f_{enc}(S^m)=X^n$ represented by a conditional distribution $P(X^n|S^m)$. 
We consider a BSC characterized by the joint conditional distribution $P(Y_{B}^n,Y_{E}^n|X^n)$, $Y_{B,i}, Y_{E,i} \in \mathcal{X}$. The noisy codeword received by Bob is decoded as $f_{dec}(Y_{B}^n)=\hat{S}^m$, and
Eve receives its own noisy observation $Y_{E}^n$.

We model the joint distribution of $T,S^m,X^n,Y_{B}^n,\hat{S}^m$, i.e., the random variables (r.v.'s) for the sensitive information, source signal, transmitted codeword, noisy codeword received by Bob, and his reconstruction, respectively, using the following graphical model $T-S^m \rightarrow X^n \rightarrow Y_{B}^n \rightarrow \hat{S}^m$ as:

\vspace{-0.5cm}
\begin{align}
    P&(T,S^m,X^n,Y_{B}^n,\hat{S}^m)=  \nonumber \\
    &P(T,S^m)P(X^n|S^m)P(Y_{B}^n,Y_{E}^n|X^n)P(\hat{S}^m|Y_{B}^n).
\end{align}


\vspace{-0.1cm}
The two BSCs independently flip each bit in the transmitted codeword with crossover probabilities $\epsilon_B$ and $\epsilon_E$ at Bob's and Eve's channels, respectively. Hence, the joint probability of the channel can be decomposed as follows:

\vspace{-0.5cm}
\begin{align}
    &P(Y_{B}^n|X^n)=\prod_{i=1}^n \epsilon_B^{x_i \oplus {y}_{B,i}}(1-\epsilon_B)^{x_i \oplus {y}_{B,i} \oplus 1}, \\
    &P(Y_{E}^n|X^n)=\prod_{i=1}^n \epsilon_E^{x_i \oplus {y}_{E,i}}(1-\epsilon_E)^{x_i \oplus {y}_{E,i} \oplus 1},
\end{align}
where $\oplus$ represents the exclusive OR operation, and $x_i$, ${y}_{B,i}$ and ${y}_{E,i}$ are the $i^{th}$ bits of $X^n$, ${Y}^n_{E}$ and ${Y}^n_{B}$, respectively.

We formulate the optimization problem as

\vspace{-0.5cm}
\begin{align}
    & & \min \limits_{f_{enc},f_{dec}} &\mathbb{E}[d({S}^m,\hat{S}^m)] - I(S^m;{Y}^n_B) + \lambda I(T;{Y}^n_E) \nonumber \\
    & & s.t.\ \ \ \ \ 
    &T,S^m \rightarrow X^n \rightarrow Y_{B}^n \rightarrow \hat{S}^m,
    \label{eq:Problem}
\end{align}
where $\lambda$ is the tuning parameter for the privacy level. 
Here, in addition to the reconstruction distortion between $S^m$ and $\hat{S}^m$, measured by $d(\cdot,\cdot)$, we also maximize the MI between the user's data $S^m$ and the noisy codewords observed by Bob, i.e., $I(S^m;{Y}^n_B)$, for improved utility. While minimizing the distortion $\mathbb{E}[d({S}^m,\hat{S}^m)]$ improves pixel-wise data reconstruction quality, we have observed in our simulations that maximizing the MI between the source signal and Bob's channel output enhances the information flow and helps with capturing the high level features at the receiver side.

Exact calculation of the MI is difficult when the data distribution is not known. 
Hence, we approximate $I(S^m;{Y}^n_B)$ and $I(T;{Y}^n_E)$ via their variational representations \cite{BA_MI}.
Due to the intractability of the true posteriors $P(S^m|{Y}^n_B)$ and $P(T|{Y}^n_E)$, we use their amortized variational approximations $f_{enc}(Y_{B}^n)=P(\hat{S}^m|Y_{B}^n)$ and $f_{eve}(Y_{E}^n)=P(\hat{T}|Y_{E}^n)$, respectively. Here, we assume that the eavesdropper tries to predict the sensitive information $T$ as $f_{eve}(Y_{E}^n)=\hat{T}$. We can write $I(S^m;Y_{B}^n)$ as follows:

\vspace{-0.4cm}
\begin{align}
    & I(S^m;{Y}^n_B)& = H(S^m)- & H(S^m|{Y}^n_B)  \label{eqn:DefMI}\\
    & & = H(S^m) + & \text{D}(P(S^m|{Y}^n_B)||f_{dec}({Y}^n_B)) \nonumber \\ 
    & &  &  \hspace{1.2cm} + \mathbb{E}[\log f_{dec}({Y}^n_B)] \label{eqn:identityMI} \\
    & & \geq H(S^m)+ & \max \limits_{f_{dec}}\mathbb{E}[\log f_{dec}({Y}^n_B)], \label{eqn:BAMI}
\end{align}
where D$(\cdot\|\cdot)$ denotes the KL divergence, $H(S^m)$ is constant, (\ref{eqn:DefMI}) follows from the definition of MI, (\ref{eqn:identityMI}) holds for any distribution $f_{dec}({Y}^n_B)$ over $S^m$ given the values in ${Y}^n_B$. Finally, (\ref{eqn:BAMI}) follows from the fact that maximum is attained when the decoder is optimum, i.e., $f_{dec}({Y}^n_B)=P(S^m|{Y}^n_B)$.
Likewise, the information leakage to the eavesdropper becomes

\vspace{-0.6cm}
\begin{align}
    & I(T;{Y}^n_E)& = H(T) - &H(T|{Y}^n_E)  \label{eqn:DefMI2}\\
    & & = H(T)+ & \text{D}(P(T|{Y}^n_E)||f_{eve}({Y}^n_E)) \nonumber \\ 
    & & & \hspace{1cm} + \mathbb{E}[\log f_{eve}({Y}^n_E)] \label{eqn:identityMI2} \\
    & & \geq H(T)+ &\max \limits_{f_{eve}}\mathbb{E}[\log f_{eve}({Y}^n_E)], \label{eqn:BAMI2}
\end{align}
where $H(T)$ is a constant term, (\ref{eqn:DefMI2}), (\ref{eqn:identityMI2}) and (\ref{eqn:BAMI2}) follow similarly to (\ref{eqn:DefMI}), (\ref{eqn:identityMI}) and (\ref{eqn:BAMI}), respectively. Here,
(\ref{eqn:BAMI}) is attained when the decoder is optimum since we maximize $I(S^m;{Y}^n_B)$ in our objective. However, (\ref{eqn:BAMI2}) is not attained even if the classifier representing the eavesdropper is optimum, because we minimize $I(T;{Y}^n_E)$ in the objective. This is due to intractability of representing $I(T;{Y}^n_E)$ with an upper-bound \cite{poole2019variational}.
On the other hand, our numerical results indicate
that although we do not optimize exact bounds for MI terms, in
practice our model still learns an effective PUT.

\vspace{-0.3cm}
\subsection{PARALLEL-CHANNEL SCENARIO}
\begin{figure}[pt]
\centering
\includegraphics[width=8.6cm]{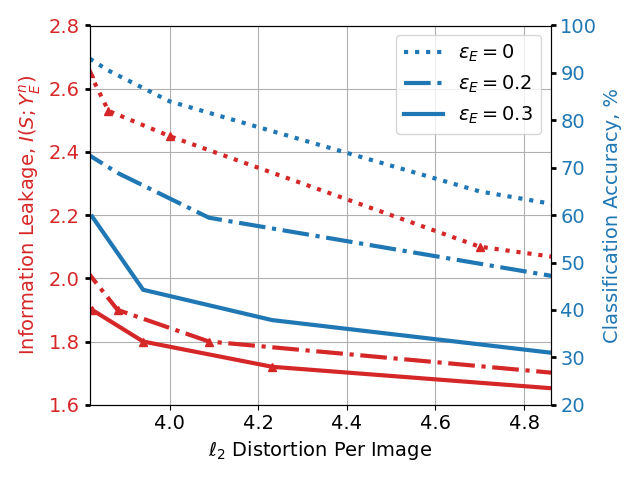}
\caption{PUT curve of our privacy-aware JSCC mechanism for $\epsilon_B=0.1$ and $\epsilon_E=\{0,0.2,0.3\}$.} 
\label{fig:MIplot}
\end{figure}

\vspace{-0.1cm}
In this section, we assume the codewords are transmitted over parallel channels with different noise levels, e.g., due to OFDM. 
Our setting represents the scenario in which the transmitter divides the total available bandwidth into non-overlapping bands carrying separate portions of the data. Each of the parallel bands face a different noise level for both the receiver and the eavesdropper, i.e., $(\epsilon_{B_i},\epsilon_{E_i})$.
For instance, in a three-channel scenario with equal bandwidths $n/3$, crossover probabilities $\epsilon_B=\{\epsilon_{B_1}, \epsilon_{B_2}, \epsilon_{B_3}\}$ and $\epsilon_E=\{\epsilon_{E_1}, \epsilon_{E_2}, \epsilon_{E_3}\}$, channel probabilities can be written as

\vspace{-0.55cm}
\begin{align}
    &P(Y_B^n|X^n)=\prod_{i=1}^{\frac{n}{3}} \epsilon_{B_1}^{x_i \oplus y_{B,i}}(1-\epsilon_{B_1})^{x_i \oplus y_{B,i} \oplus 1} \label{eq:MultichannelProbR}
    \\
    &\times
    \hspace{-0.4cm}
    \prod_{j=\frac{n}{3}+1}^{\frac{2n}{3}} 
    \hspace{-0.3cm}
    \epsilon_{B_2}^{x_j \oplus y_{B,j}} \hspace{-0.06cm} ( \hspace{-0.06cm} 1 \hspace{-0.1cm} - \hspace{-0.06cm} \epsilon_{B_2} \hspace{-0.06cm} )^{x_j \oplus y_{B,j} \oplus 1} 
    \hspace{-0.42cm}
    \prod_{k=\frac{2n}{3}+1}^{n}
    \hspace{-0.41cm}
    \epsilon_{B_3}^{x_k \oplus y_{B,k}} \hspace{-0.06cm} ( \hspace{-0.06cm} 1 \hspace{-0.1cm} - \hspace{-0.06cm} \epsilon_{B_3} \hspace{-0.06cm} )^{x_k \oplus y_{B,k} \oplus 1} \nonumber
\end{align}
for the receiver, and as follows for the eavesdropper:

\vspace{-0.6cm}
\begin{align}
    &P(Y_E^n|X^n)=\prod_{i=1}^{\frac{n}{3}} \epsilon_{E_1}^{x_i \oplus y_{E,i}}(1-\epsilon_{E_1})^{x_i \oplus y_{E,i} \oplus 1} \label{eq:MultichannelProbE}
    \\
    &\times
    \hspace{-0.4cm}
    \prod_{j=\frac{n}{3}+1}^{\frac{2n}{3}} 
    \hspace{-0.3cm}
    \epsilon_{E_2}^{x_j \oplus y_{E,j}} \hspace{-0.06cm} ( \hspace{-0.06cm} 1 \hspace{-0.1cm} - \hspace{-0.06cm} \epsilon_{E_2} \hspace{-0.06cm} )^{x_j \oplus y_{E,j} \oplus 1} 
    \hspace{-0.4cm}
    \prod_{k=\frac{2n}{3}+1}^{n}
    \hspace{-0.4cm}
    \epsilon_{E_3}^{x_k \oplus y_{E,k}} \hspace{-0.06cm} ( \hspace{-0.06cm} 1 \hspace{-0.1cm} - \hspace{-0.06cm} \epsilon_{E_3} \hspace{-0.06cm} )^{x_k \oplus y_{E,k} \oplus 1} \hspace{-0.1cm}. \nonumber
\end{align}

\vspace{-0.25cm}
We solve (\ref{eq:Problem}) using the channel probabilities (\ref{eq:MultichannelProbR}) and (\ref{eq:MultichannelProbE}). We want our solution for (\ref{eq:Problem}) to control the transmission through the channels such that the sensitive information $T$ is transmitted over the channels in which Eve experiences high noise, while the rest of the source is transmitted over the channels Bob experiences low noise, independent of Eve's channel. We numerically verify that the proposed VAE-based encoder indeed satisfies these expectations.

\vspace{-0.45cm}
\section{SIMULATION RESULTS}
\vspace{-0.3cm}
\label{sec:Sims}
We consider the wiretap channel in Fig. \ref{fig:wiretapscenario}, where the encoder and decoder at Alice and Bob are represented by a VAE, while Eve employs a classifier. For the encoder-decoder pair, we employed the network structure ``NECST'' proposed in \cite{necst}. We designed our privacy aware JSCC network by incorporating our classifier based eavesdropper in NECST.
We used colored MNIST handwritten digits as $S^m$ for $m=32\times32$ pixels, and color and thickness of the digits as the sensitive information $T \in \mathcal{T}=\{(R,0),$ $(R,1),(R,2),(G,0),(G,1),(G,2),(B,0),(B,1),(B,2)\}$, where $R$, $G$ and $B$ denote red, green and blue colors, while $0$, $1$ and $2$ represent thin, medium and thick digits, respectively. We set the total channel bandwidth to $n=200$ bits. We present MI using mutual information neural estimator (MINE) \cite{poole2019variational}, which is a realistic approximation of MI. Due to its high computational complexity, we only
used \cite{poole2019variational} for calculating the information leakage in the test time.

\vspace{-0.4cm}
\subsection{Single Channel}

\vspace{-0.2cm}
We first consider a single channel scenario. In Fig. \ref{fig:MIplot}, information leakage $I(T;Y^n_E)$ and Eve's classification accuracy are shown with respect to $\ell_2$ distortion per image. Dotted, dashed and straight lines represent the cases with $\epsilon_E=0$, $\epsilon_E=0.2$ and $\epsilon_E=0.3$, respectively, while we have $\epsilon_B=0.1$ for all cases. Data points are taken at $\lambda=\{0, 5, 10, 20\}$.
Fig. \ref{fig:MIplot} shows that the information leakage about the sensitive information decreases as the image distortion increases, which is expected due to the PUT. This is the case even when there is no noise in Eve's channel, i.e., $\epsilon_E=0$, since minimizing the MI  between $T$ and $Y^n_E$ provides a certain level of privacy. Moreover, noisier eavesdropper channel leaks less information at the same level of distortion. Similar trend can be seen for Eve's accuracy. In Fig. \ref{fig:MIplot}, we also observe that a MI gap as small as $0.06$ corresponds to $20\%$ accuracy gap between $\epsilon_E=0.2$ and $\epsilon_E=0.3$ cases. 

\begin{figure}[pt]
\centering
\subfloat[$\lambda=0$]{\includegraphics[width=8.6cm]{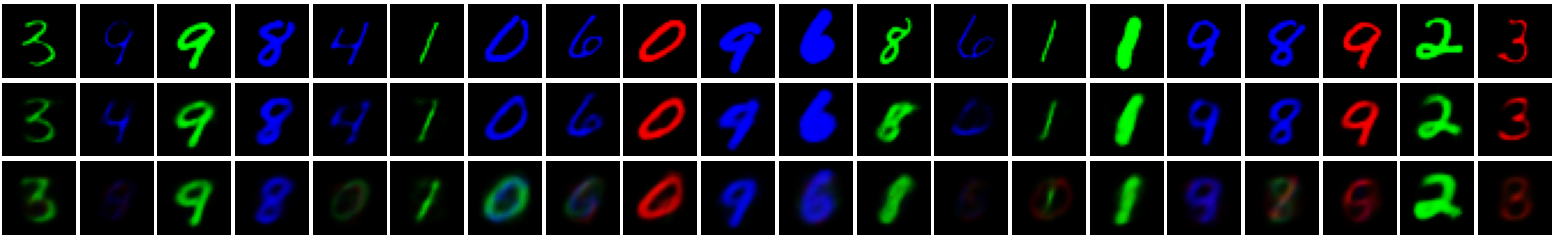}}
\vfill
\vspace{-0.4cm}
\subfloat[$\lambda=20$]{\includegraphics[width=8.6cm]{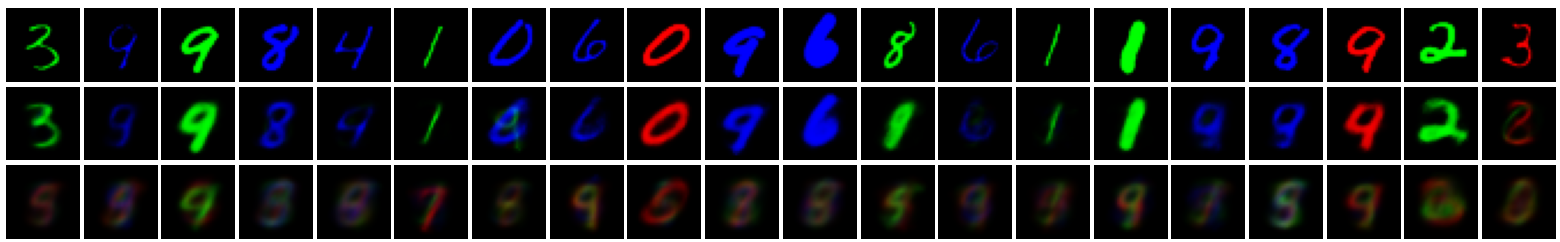}}

\vspace{-0.15cm}
\caption{Original images and their reconstructions by Bob and Eve from top to bottom, respectively, for $\epsilon_B=0.1$, $\epsilon_E=0.3$.} 
\label{fig:SinglechannleReconst}
\end{figure}

For illustration purposes, we trained an additional decoder on the noisy bits received by Eve ($Y^n_E$) with the same structure as Bob's decoder. Fig. \ref{fig:SinglechannleReconst} depicts the original images, reconstructed images by Bob and Eve, respectively, from top to bottom. We can see that in the absence of privacy ($\lambda=0)$, both Bob and Eve can reconstruct the images rather accurately, while, thanks to the employed JSCC approach, Bob's better channel allows it to have better fidelity.  On the other hand, when privacy is imposed ($\lambda=20)$, we can see that Eve cannot recover neither the colour nor the thickness information. On the other hand, we can see that this information is available to Bob; and hence, it has been successfully hidden from Eve while being available in the transmitted signal.



\begin{table}[pb]
\caption{Information leakage and Eve's classification accuracy for the sensitive r.v. $\hat{T}$ and individual sensitive attributes at each channel for $\lambda=10$}
\label{tab:smooth_results}
\centering
\setlength{\tabcolsep}{3pt}
\begin{tabular}{ lcccc }
 \toprule
 \textbf{Channels}  & 
 \textbf{Ch1} & \textbf{Ch2} & \textbf{Ch3} & \textbf{Ch4}     \\
\midrule
\textbf{$I(T;Y_E^n)$}      
& $1.0836$ & $1.5703$ & $0.0689$ & $0.6411$ \\      \midrule
\textbf{Accuracy, $T$}  
 & $13.65\%$ & $31.85\%$ & $16.15\%$ & $19.6\%$ \\
 \midrule
\textbf{Accuracy, Color}  
 & $34.5\%$ & $62.35\%$ & $41.2\%$ & $45.25\%$ \\
 \midrule
\textbf{Accuracy, Thickness}  
 & $35.75\%$ & $48.65\%$ & $38.05\%$ & $38.55\%$ \\
\bottomrule 
\end{tabular}
\end{table}

\vspace{-0.4cm}
\subsection{Parallel Channels}

\vspace{-0.2cm}
Next, we consider a parallel-channel scenario, where the signal is transmitted over multiple channels with different noise levels. We use $4$ parallel channels each with a bandwidth of $n/4=50$ bits. Error probability pairs for Bob's and Eve's channels are set as $(\epsilon_B,\epsilon_E)=\{\text{Ch1}:(0.1,0.1), \text{Ch2}:(0.001,0.2), \text{Ch3}:(0.2,0.001),  \text{Ch4}:(0.001,0.001)\}$.
Table \ref{tab:smooth_results} shows the information leakage, Eve's classification accuracy on $T$, and separately on the sensitive attributes \textit{color} and \textit{thickness} for each channel. \textit{Accuracy, Color} and \textit{Accuracy, Thickness} are calculated as the success of the classifications for only the \textit{color} and only the \textit{thickness}, respectively. Our privacy-aware generative network obtains the PUT by minimizing the information leakage of the sensitive attributes and the distortion. This leads to smaller information leakage at the best quality channel of Eve, i.e., Ch2, and larger at the worst one, i.e., Ch3. Eve's classification accuracy of $T$, and individual color and thickness attributes, are the highest for Ch2 and lowest for Ch1. We observed that Ch1 accuracy is low because the classifier is confused between \textit{blue} and \textit{green}, as well as the \textit{medium} and \textit{thick}, but still has high accuracy for \textit{red} and \textit{thin} attributes. On the other hand, Ch3 has low accuracy for all the attributes. This leads to the difference between the leakage and accuracy results for Ch1 and Ch3.

\begin{figure}[pt]
\centering
\includegraphics[width=8.6cm]{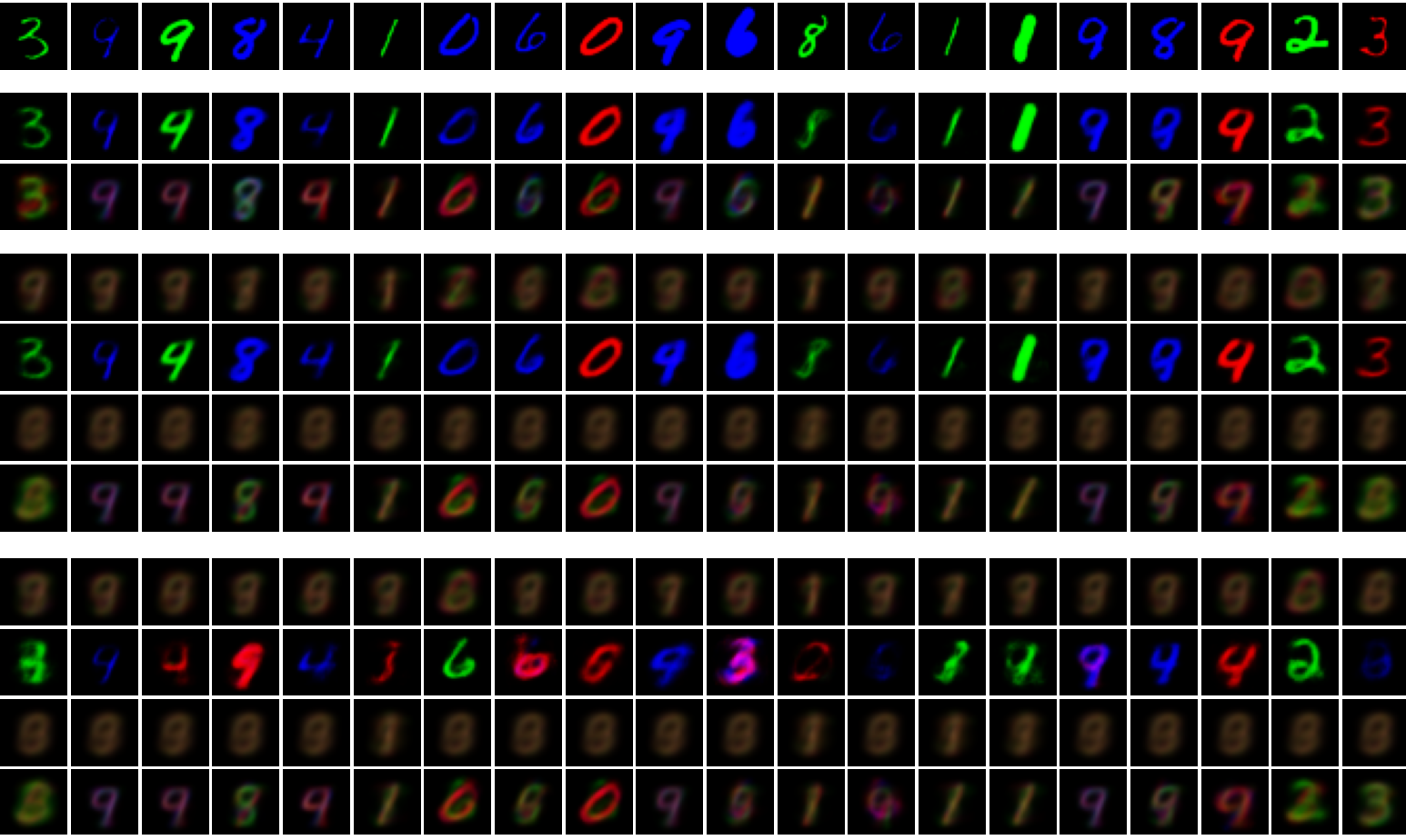}

\vspace{-0.15cm}
\caption{Original images and reconstructions by Bob, Eve, Bob's individual channels (Ch1-4), and Eve's channels (Ch1-4), respectively, from top to bottom, for $\lambda=10$.} 
\label{fig:MultichannelReconstruct}
\end{figure}

In Fig. \ref{fig:MultichannelReconstruct}, we show the original and reconstructed images by Bob, Eve, Ch1 to Ch4 of Bob, and Ch1 to Ch4 of Eve, respectively, from top to bottom. First three rows show similar results with the single channel case, i.e., Eve is confused about the color and thickness of the digits, while Bob can reconstruct at high quality. Moreover, Ch1 and Ch3 do not have meaningful reconstructions for either Bob or Eve. This is because Eve faces less noise in these channels, which might lead to larger leakage. Hence, our network minimizes the information flow through these channels. Ch2, on the other hand, carries more information than Ch4 since it can better hide the sensitive attributes from Eve while maximizing the information transmission for Bob.

\vspace{-0.5cm}
\section{CONCLUSION}
\vspace{-0.35cm}
We proposed a VAE-based privacy-aware communication scheme over a wireless wiretap channel. In our simulation results, we showed that our end-to-end learning approach provides minimally distorted source transmission with maximum channel capacity while minimizing the information leakage about sensitive information to an eavesdropper. We also showed that our approach balances the information flow in a parallel-channel scenario such that the PUT is obtained according to the receiver's and eavesdropper's channel noises.



\newpage

\newpage
\bibliographystyle{IEEEbib}
\bibliography{refs}

\end{document}